\documentclass[10pt]{iopart}
\usepackage{iopams} 
\usepackage[english]{babel}

\usepackage[normalem]{ulem}

\usepackage{bm}
\usepackage{graphics}
\usepackage{graphicx}

\usepackage[utf8x]{inputenc}
\usepackage[T1]{fontenc}
\usepackage{ae,aecompl}

\usepackage[usenames,dvipsnames]{xcolor}

\usepackage{color}
\definecolor{darkgreen}{rgb}{0,0.6,0}
\definecolor{darkblue}{rgb}{0,0,0.6}
\definecolor{darkred}{rgb}{0.6,0,0}
\definecolor{darkpurple}{rgb}{0.5,0,0.5}

\usepackage{hyperref}

\hypersetup{
bookmarksopen=true,
pdftitle=Building a path-integral calculus: a covariant discretization approach,
pdfauthor=L F Cugliandolo V Lecomte F van Wijland, 
pdftoolbar=false, % toolbar hidden
pdfstartview={FitH},		% fits the width of the page to the window
pdfmenubar=true,			%menubar shown
pdfhighlight=/O,			%effect of clicking on a link
colorlinks=true,			%couleurs sur les liens hypertextes
urlcolor=darkblue,
citecolor=darkblue,		%couleur des liens pour les citations
linkcolor=darkpurple	%darkred %couleur des liens hypertextes internes
}

\newenvironment{align}
 {\begin{eqnarray}}
 {\end{eqnarray}\ignorespacesafterend}

\definecolor{linkcolor}{rgb}{0,0,0.6} 
\definecolor{brown}{RGB}{112,10,10}
\definecolor{darkred}{RGB}{199,23,28}

\newcommand{\dd}{\text{d}}

\newcommand{\ee}{\text{e}}

\newcommand{\eeref}[1]{Eq.$\:$(\ref{#1})}
\newcommand{\Tfg}{\mathbb T_{\!f\!,g}}
\newcommand{\TFG}{\mathbb T_{\!F\!,G}}

\newcommand{\tf}{t_{\text{f}}}
\newcommand{\st}{{\textnormal{\tiny{S}}}}
\newcommand{\deltaS}{\delta\mkern-1.5mu S}

\newcommand{\xdt}{x_{\mkern-1.mu\Delta t}}

\newcommand{\betag}{\mkern-1.mu\beta_{g}\mkern-1.4mu}
\newcommand{\betaG}{\beta_{G}\mkern-1.4mu}
\newcommand{\PP}{\mathbb P}

\renewcommand{\phi}{\varphi}
\newcommand{\NN}{\mathcal N}

\providecommand{\tfrac}[2]{\textnormal{$\frac{#1}{#2}$}}
\providecommand{\eqref}[1]{\eref{#1}}
\providecommand{\text}[1]{{\rm{#1}}}

\begin{document}

\title{Building a path-integral calculus: a covariant discretization approach}

\author{Leticia F.~Cugliandolo$^{1}$, Vivien~Lecomte$^{2}$ and Frédéric~van~Wijland$^{3}$}

\address{$^{1}$
Laboratoire de Physique Théorique et Hautes Énergies, Sorbonne Université,  UMR 7589 CNRS, 
4 Place Jussieu Tour 13 5\`eme \'etage, 75252 Paris Cedex 05, France
\smallskip
}

\address{$^{2}$
Laboratoire Interdisciplinaire de Physique, Université Grenoble Alpes, CNRS, LIPhy, F-38000 Grenoble, France
\smallskip
}

\address{$^{3}$
Laboratoire Matière et Systèmes Complexes, Université Paris Diderot, UMR 7057 CNRS, 10 rue Alice Domon et Léonie Duquet, 75205 Paris Cedex 13, France
\smallskip
}

% \significancestatement{
% Path integrals are ubiquitous in theoretical physics because they allow for a vivid and technically efficient description of fluctuating objects, whether they are quantum or stochastic. However, it has been known almost since the beginning that path integrals do not lend themselves to the same intuitive rules as those used in ordinary differential calculus, based on derivatives and changes of variables. We identify what the hitherto overlooked missing ingredients are in constructing such well-behaved path integrals. With our construction, the validity of which we mathematically establish, one can thoughtlessly work with functional integrals over paths as one likes working with regular integrals of functions over the real axis. This opens the road for a sound path-integral based calculus.}

%\correspondingauthor{\vspace*{-2mm}\textsuperscript{2}To whom correspondence should be addressed. E-mail: vivien.lecomte@univ-grenoble-alpes.fr}

\ead{\\leticia@lpthe.jussieu.fr\\ vivien.lecomte@univ-grenoble-alpes.fr\\frederic.van-wijland@univ-paris-diderot.fr}

\begin{abstract}
Path integrals are a central tool when it comes to describing quantum or thermal fluctuations of particles or fields. Their success dates back to Feynman who showed how to use them within the 
framework of quantum mechanics. Since then, path integrals have pervaded all areas of physics where fluctuation effects, quantum and/or thermal, are of paramount importance. Their 
appeal is based on the fact that one converts a problem formulated in terms of operators {into} 
one {of} sampling classical paths with a given weight. Path integrals are the mirror image of our conventional Riemann integrals, with functions replacing the real numbers one usually sums 
over. However, unlike conventional integrals, path integration suffers a serious drawback: in general, one
cannot make non-linear changes of variables
without committing an error of some sort. {Thus,} 
no path-integral based calculus is possible. Here we identify which are  the deep mathematical reasons causing this important caveat, and we come up 
with
cures for systems described by one degree of freedom. Our main result is a construction of 
path integration free of this longstanding problem, through a direct time-discretization procedure. 
\end{abstract}


\bigskip

%_____________________________________________________________________________________________________

\newpage
\begin{small}
	\tableofcontents
\end{small}

\maketitle

\section{Introduction}
\label{sec:introduction}

Though the  {notion} 
of path integration can be traced back to Wiener~\cite{wiener_differential-space_1923,wiener_average_1924}, it is fair to credit Feynman~\cite{feynman_space-time_1948} for 
making {path integrals one of the} daily tools of theoretical {physics}. 
The idea is to express the transition amplitude {of a particle} between two 
states as an integral over all possible trajectories {between these states} with an appropriate weight for each of them. 
After such a formulation of quantum mechanics was proposed, path integrals turned out to provide a set of methods 
that are now ubiquitous in Physics (see~\cite{chaichian_stochastic_2001,zinn-justin_quantum_2002,kleinert_path_2009} for reviews)
and they have become the language of choice for quantum field theory. 
But path integrals reach out well beyond quantum physics and they are also a versatile instrument
{to study} stochastic processes. Beyond Wiener's original formulation of Brownian motion, Onsager and Machlup~\cite{onsager_fluctuations_1953,machlup_fluctuations_1953II}, followed by  Janssen~\cite{Janssen1976,Janssen1979}, and De Dominicis~\cite{dominicis_techniques_1976,DeDominicis1978} (based on the operator formulation of Martin, Siggia and 
Rose~\cite{Martin1973}), have contributed to establish path integrals as a useful tool, on equal footing with the
Fokker--Planck and Langevin equations. 
The gist of the mathematical difficulty is to manipulate signals that are nowhere differentiable.
Interestingly, mathematicians have mostly stayed a safe distance away from path integrals. Indeed, it has been known for many years that path integrals cannot be manipulated without extra caution in a vast category of problems. These problems, in the stochastic language, 
involve the notion of multiplicative noise (that we describe in detail below), and their counterpart in the quantum world has to do with  quantization on curved spaces~\cite{bastianelli_path_2006}. 

The late seventies witnessed an important step toward the understanding of the subtleties of path integrals: the authors of \cite{horsthemke_onsager-machlup_1975,graham_path_1977,graham_covariant_1977,weiss_operator_1978,kerler_definition_1978,deininghaus_nonlinear_1979} found how to formulate path integrals in terms of smooth (differentiable) functions.
By construction, their formulation does not offer a direct interpretation in terms of the weight of the physical trajectories, which are non-differentiable.
%
% Yet, a path integral acquires a definite meaning only as the continuum limit of a discretized expression~\cite{fujiwara_rigorous_2017} and this was not achieved in these references.
%
The goal of this article is to come up with the missing link: we
construct path integrals for non-differentiable stochastic and/or quantum trajectories, free of any mathematical {hitch}, 
by a direct time-discretization procedure which endows them with a well-defined mathematical meaning, consistent with differential calculus.

\section{Result and outline}
%
% we announce here the mathematical statement of this article:
%
Consider a system described by a single degree of freedom $x(t)$ with noisy dynamics  
(\emph{i.e.},~subjected to a random force).
We give an unambiguous definition of the probability density $\mathbb P$
of a path $[x(t)]_{0\leq t\leq \tf}$ in a form that is \emph{covariant} under any change of variables $u(t)=U(x(t))$.
Namely, denoting by $x_k$ and $u_k$ the sequences of values that the paths $x(t)$ and $u(t)$ take at discrete times indexed by an integer $k$,
the probability density $\mathbb P$ of such sequences satisfies
\begin{equation}
\prod_{k=0}^N dx_k \
\mathbb P_X[\{x_\ell\}]
= 
\prod_{k=0}^N du_k 
 \ \mathbb P_U[\{u_\ell\}]
 \; , 
\label{eq:goal}
\end{equation}
with $\mathbb P_X$ and $\mathbb P_U$ taking {\it the same functional form} for the processes 
$\{ x_\ell \}$ and $\{ u_\ell=U(x_\ell)\}$. In these expressions,  $U$ is an arbitrary invertible differentiable function and $N$ is the number of time steps in which the time window $[0,\tf]$ is divided.
The precise definitions of all the entities involved in the relation~\eref{eq:goal} will be given in the central part of this paper (Section~\ref{sec:probtraj}).

%\textcolor{red}{Is this expressed in a precise enough form? Check}

The continuous-time limit of $\mathbb P[x]$ reads $\mathcal N[x] \ee^{-S[x]}$ where both the `action'~$S[x]$ and the `normalization factor'~$\mathcal N[x]$ are covariant: 
in the Lagrangian writing $S[x]=\int_0^{\tf\!}\dd t\, \mathcal L(x,\dot x)$,
% Feu! Chatterton
%
switching between $x(t)$ and $u(t)$  merely amounts to applying the chain rule $\dot u(t)=\dot x(t)\,U'(x(t))$.
We emphasize that, from our theoretical physicist's point of view, $\mathbb P[x]$ acquires the meaning of the probability of a path only when a discretized version is given and
such a discretization issue is not a mathematical detail:
continuous-time writings of $\mathcal N[x]$ and $S[x]$  do not allow one to identify without 
ambiguity the probability of a path\footnote{
We underline here an important cultural difference with a mathematician's viewpoint which would consist in defining a path-integral action directly in continuous time, following Wiener~\cite{wiener_differential-space_1923,wiener_average_1924} (and others~\cite{takahashi_probability_1981,hara_lagrangian_1996,capitaine_onsager-machlup_2000} for multiplicative processes). Our point of view is different: we prefer to keep an underlying time discretization with infinitesimal time step $\Delta t$ which allows us to control the rest in powers of $\Delta t$ when manipulating the action, evaluated on non-differentiable paths.
From a mathematician's viewpoint, we are interested in the probability density of the events $\{x(t_k)=x_k\}$.
}.
 The discretization scheme that we present in this work is compatible with the covariance relation~\eref{eq:goal} and solves the long-standing problem 
of building a well-defined path probability that is consistent with differential calculus. 

In what follows, we construct our path integral by carefully manipulating non-differentiable trajectories, directly from a Langevin equation.
The latter suffers from ambiguities that only a discretized formulation can waive, and we thus begin in Section~\ref{sec:stocproc} with a review of discretization issues in Langevin equations. With this settled, we present in Section~\ref{sec:probtraj} the main outcome of our paper, a path probability (that includes a carefully defined normalization factor) that allows one to use the standard rule of calculus inside the action when changing variables, even in the time-discrete formulation~\eref{eq:goal} and for non-differentiable trajectories.
Constructing the actual time-discrete path probability requires to focus on hitherto overlooked contributions in slicing up time-evolution, but also to resort to a new adaptive slicing of time. 
It amounts to identifying the correct discretization of the integral $S[x]=\int_0^{\tf\!}\dd t\, \mathcal L(x,\dot x)$, an issue that goes well beyond the usual Itō-Stratonovich dilemma,
and thus enforces us to implement a generalization of the standard stochastic integral.
In Section~\ref{sec:comparotherconstrc} we compare our result and our construction to other path-integral formulations.
In Section~\ref{sec:msr}, we then show how to transpose our construction to the so-called Martin--Siggia--Rose--Janssen--De~Dominicis~\cite{Janssen1976,Janssen1979,dominicis_techniques_1976,DeDominicis1978,Martin1973} (MSRJD) 
path-integral representation of the path probability, 
that provides the Hamiltonian counterpart of the former Lagrangian formulation [the action $S[\hat x,x]$ now depending on a `response variable'  $\hat x(t)$ 
conjugate to $x(t)$]. We finally provide in Section~\ref{sec:summary} our conclusion and outlook.

\section{Stochastic processes}
\label{sec:stocproc}

For concreteness,
we focus on the problem of a point-like particle moving in a one dimensional space.

\subsection{Langevin's Langevin equation}
Langevin introduced the celebrated equation that goes under his name to describe Brownian motion~\cite{LangevinBrownian}. %{lemons_paul_1997}.
%\textcolor{red}{not sure we need to give references here, maybe just Langevin's paper?}
His idea was to start from Newton's equation $m\dot v(t) = F(t)$ for the motion of the large particle with mass $m$ and velocity $v$,
and to mimic the effect of its 
contact with the embedding liquid through a phenomenological force $F(t)$ made of two terms: a dissipative contribution, $-\gamma v(t)$, 
and a time-dependent random one,~$\eta(t)$. With this simple choice for the former and adopting adequate statistical properties for the latter, 
he represented the observed erratic motion of the particle, and understood the behavior of varied experimentally averaged observables,
constructed in his formalism as averages [denoted $\langle\,\cdot\,\rangle$] over the noise. Importantly enough, 
he assumed that the random force was Gaussian distributed at each instant, had zero mean, $\langle \eta(t)\rangle=0$, and was 
Dirac-delta correlated in time, $\langle \eta(t)\eta(t')\rangle=2D\delta(t-t')$, assuming a strong separation of time-scales between 
the one of the motion of the Brownian particle and the ones 
typical of the motion of the constituents of the ``bath''.  Such a ``thermal noise'' $\eta(t)$ is termed {\it Gaussian white noise}. 
Denoting by $k_{\text{B}}$ the Boltzmann constant and by $T$ the ambient temperature, 
the parameter $D$ is fixed to $\gamma k_{\text{B}}T$ in order
to ensure kinetic energy equipartition.
In the so-called overdamped limit one studies time-scales 
that are much longer than $m/\gamma$, neglecting inertia compared to the effect of other forces, 
and focuses on the particles's position $x(t)$ that is ruled by 
$\dot x (t)= f(x(t)) +\eta(t)$. 
In this notation  the friction coefficient $\gamma$ was absorbed in 
a redefinition of time,
and a term $f(x(t))$, proportional to an external force, was added to describe more general physical situations. 

\subsection{Reductionism: other Langevin equations}
Stochastic equations of the Langevin kind have later been derived for the dynamics of other degrees of
freedom than the position, or of fluctuating order parameters in (even originally quantum) systems, after a \emph{model reduction} that amounts to integrating over a large number of degrees of freedom
in an interacting system, keeping only a few representative ones. The range of applicability of Langevin equations therefore
became much wider than originally expected~\cite{kampen_stochastic_2007,gardiner_handbook_1994}. 
A large separation of time-scales is also usually advocated to 
claim that Gaussian white noise is a reasonable choice and, furthermore, the overdamped limit is also often justified.

\subsection{Multiplicative noise}
In many cases of practical interest the noise is not additive as in Langevin's original 
proposal but appears multiplied by a function of the variable of interest,
\begin{equation}\label{eq:eqLangevin}
\dot{x}(t)=f(x(t))+g(x(t))\eta(t)
\; , 
\end{equation}
still with $\langle \eta(t)\eta(t')\rangle=2D\delta(t-t')$. 
Such a multiplicative noise is involved in a flurry of physical problems ranging from soft matter (\emph{e.g.}~diffusion in microfluidic devices~\cite{grun_thin-film_2006}), to condensed 
matter (\emph{e.g.}~super paramagnets~\cite{genovese_mesoscopic_1998,birner_critical_2002}) or
even inflational cosmology~\cite{matacz_new_1997,vennin_correlation_2015}. It appears in other areas {of science}
in which Langevin equations are present ({\it e.g.}~Black--Scholes equation for option pricing~\cite{black_pricing_1973}). Quantization on curved spaces (\emph{e.g.}~{a} particle on 
a sphere~\cite{carinena_quantum_2011,carinena_quantum_2012} or more generic 
manifolds~\cite{dewitt_dynamical_1957,mclaughlin_path_1971,gervais_point_1976,grosche_path_1987}) pertains to the same mathematical class of problems, 
even though their physical motivation has a different origin.
Connections between thermal and quantum noises were noted by Nelson~\cite{nelson_quantum_1985}, 
and it is therefore no surprise that our discussion addresses both class of 
problems.
% {simultaneously}. 

\subsection{Discretization}
Langevin defined his equation and performed calculations in a continuous-time setting. 
However, an overdamped multiplicative Langevin equation such as~\eref{eq:eqLangevin} acquires a well-defined meaning only if 
a discretization scheme is chosen.
We adopt here the physicist's description where a time-discrete version of~(\ref{eq:eqLangevin}) is made explicit. 
%
%Indeed, a differential equation is just the limit of a finite difference equation in which the steps of the independent variable 
%are taken to zero in a convenient way. 
%[V: la phrase me semble un peu trop générale pour être assez précise en ce qui concerne la nécessité de la discrétisation]
%
Controlling the zero time-step limit in a careful way is crucial when dealing with
stochastic equations because $x(t)$ is not a differentiable function. 
To address this issue with the appropriate rigor, 
mathematicians have developed the field of stochastic calculus (see for instance~\cite{karatzas2012brownian,oksendal_stochastic_2013} for reviews); thus, they often use the continuous-time Wiener measure as a reference to define other structures of interest, but we do not follow this approach here because our interest goes to explicit trajectory weights.

\begin{figure}[t]
  \centering\quad\includegraphics[width=.77\columnwidth]{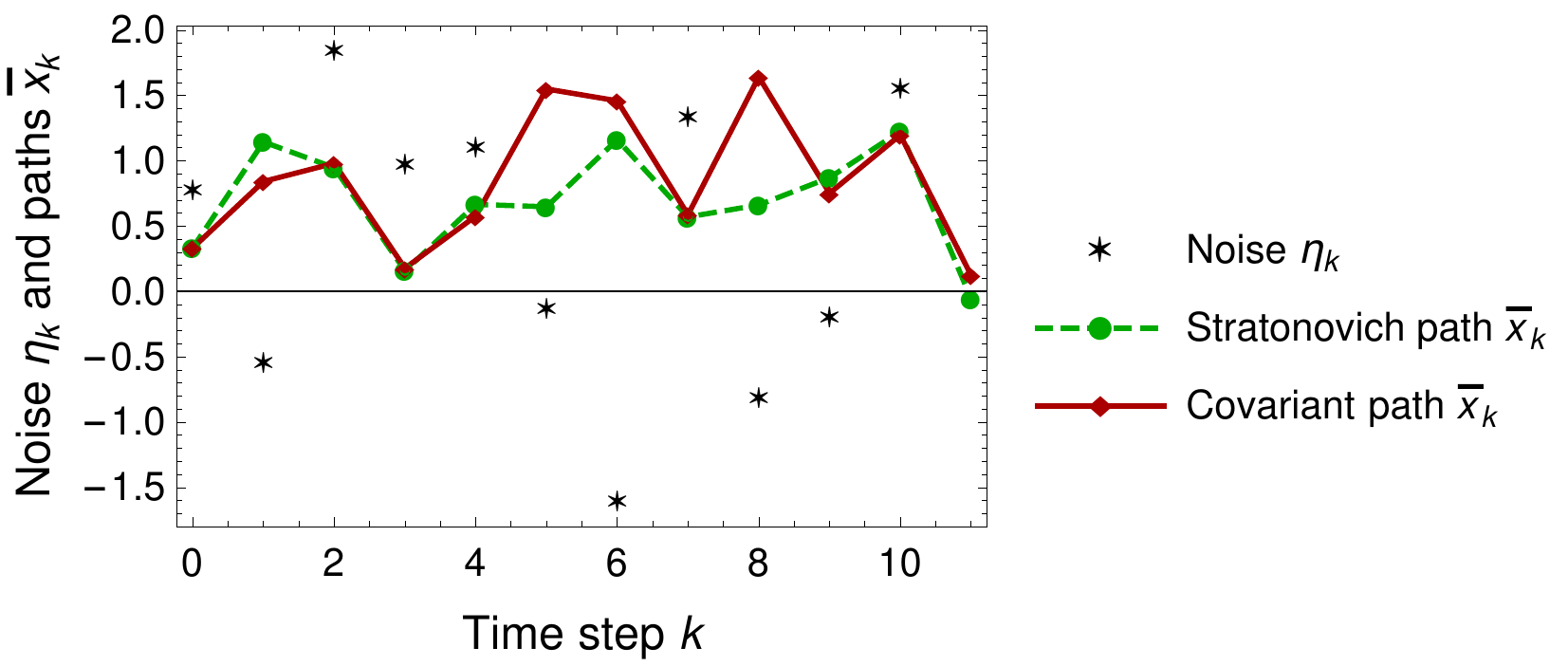} \vspace*{-3mm}
  \caption{Comparison between Langevin paths discretized {\`a la} Stratonovich (circles, 
  dashed line) and with the covariant rule (diamonds, full line), 
  according to Eqs.$\:$\eref{eq:StratoLangevin} and~\eref{eq:def-cov-discr-Tg}, 
  respectively. The noise is represented with stars. The Langevin dynamics is defined by $f(x)=1+x$, $g(x)=4x^4$, $D=1$, discretized with $\Delta t= 1/4$.
}
\label{fig:compar-discr-path}
\end{figure}

%
% For time-dependent functions and 
% equations as the ones we are dealing with here, the full time-interval going from, say $t=0$ to $t=t_f$, 
The time interval $[0,\tf]$ is divided into $N$ steps of equal duration $\Delta t$, in such a way that $t_k = k \Delta t$,
with $k=0,\dots, N$ and $N\Delta t = t_N= \tf$. The instantaneous 
noise $\eta_k = \eta(t_k)$, is drawn from the joint probability distribution function (p.d.f.)
\begin{equation}
\mathbb{P}[\{\eta_k\}] = \prod_{0\leq k<N} \sqrt{\frac{\Delta t}{4\pi D}} \;  e^{-\frac{\Delta t}{4D} \eta_k^2}
%\mathbb{P}[\{\eta_\ell\}] = \prod_{k=0}^N \frac{1}{\sqrt{4\pi D\Delta t^{-1}}} \;  e^{-\frac{\Delta t}{4D} \eta_k^2}
\; .
\label{eq:noise-pdf}
\end{equation}
A set of noises drawn from this p.d.f.~are shown in Fig.$\:$\ref{fig:compar-discr-path} with stars.
The measure over which functions of the noise are integrated over is 
%${\cal D} \eta \equiv \prod_{0\leq k<N} {\rm d}\eta_k$.
${\cal D} \eta \equiv \prod_{k=0}^{N-1} {\rm d}\eta_k$.
This p.d.f.~implies $\langle \eta_k \rangle = 0$ and 
$\langle \eta_k \eta_{k'} \rangle = 2D\delta_{kk'}/\Delta t$, making explicit that 
$\eta_k = {O}(\Delta t^{-1/2})$ at each time step.
To define the Langevin equation~\eref{eq:eqLangevin}, one now specifies the time-discrete evolution for the $x_k\equiv x(t_k)$'s (with $0\leq k\leq N$).
First, the time derivative $\dot x(t)$ evaluated at $t_k$ represents the ratio $\Delta x/\Delta t$ between the two forward increments,
 $\Delta x \equiv x_{k+1} - x_{k}$ and $\Delta t=t_{k+1}-t_k$.  
Second, to specify how to evaluate $x(t)$ on the right-hand side (r.h.s.)~of~\eref{eq:eqLangevin}, we denote by $\bar x_k$ the arguments of the functions $f$ and $g$ in the time-discrete evolution. 
In conventional stochastic calculus, $\bar x_k$ is given by a linear combination of $x_{k}$ and 
$x_{k+1}$. A dependence on the sole pre-point $\bar x_k = x_k$ is chosen in the Itō scheme and, instead,  
the mid-point dependence $\bar x_k = (x_{k}+x_{k+1})/2$ is taken in the Stratonovich one. Each form has its 
advantages and drawbacks. Within the Stratonovich convention, in the continuous-time limit, one can manipulate $x(t)$ as if it were differentiable 
but $x_{k+1}$ appears in implicit form  in the discrete equation at time $t_k$ and this is not convenient for numerical integration. 
Instead, the Itō scheme yields a recursion particularly suited to the computer generation of an individual trajectory. 
However, in contrast to \eeref{eq:eqLangevin} understood with the Stratonovich rule, one cannot manipulate $x(t)$ as if it were differentiable. This problem was addressed by mathematicians who modified the rules of calculus to be able to work with $x(t)$ in the continuous-time limit. This is the celebrated Itō's lemma~\cite{ito_stochastic_1944}. 

The continuous-time equation~\eref{eq:eqLangevin} is thus understood as a short-hand writing which acquires a well-defined meaning {\it only} through 
a limiting procedure $\Delta t\to 0$ which starts from a discrete-time evolution in which a prescription (or `discretization scheme') for $\bar x_k$ is given.
We focus on the Stratonovich choice henceforth, with the aim of building a path-integral formalism in which the standard rules of calculus could also be used.
The time-discrete evolution %of the stochastic variable $x$ 
is therefore given by 
\begin{eqnarray}
\frac{\Delta x}{\Delta t}    &    \stackrel{\st}{=}    &  f(\bar x_k) + g(\bar x_k) \eta_k
\qquad
(0\leq k<N)
\; , 
\label{eq:DeltaxDeltatStrato}
\\
\;\;\bar x_k    &    =    &   \frac{x_k+x_{k+1}}{2} =  x_k + \frac{1}{2} \Delta x
\; , 
\label{eq:StratoLangevin}
\end{eqnarray}
where $\stackrel{\st}{=}$ indicates that $\bar x_k$ is Stratonovich-discretized.
This
implies that the typical $\Delta x$ is of order $\sqrt{\Delta t}$, and not $\Delta t$, reflecting the 
well-known fact that a Brownian motion is nowhere differentiable. 
Each  choice of $\{\eta_k\}_{0\leq k < N}$
drawn from the noise p.d.f.~\eref{eq:noise-pdf} yields a trajectory 
$\{x_k\}_{0\leq k \leq N}$,
with $x_0$ drawn from 
%an arbitrary 
a
distribution $P_{\text{i}}$.
A sketch of such a trajectory is shown in Fig.~\ref{fig:compar-discr-path} with circles. 
The probability density (or `path probability') of such trajectories, $\mathbb{P}_X[\{x_k\}]$,  will be the object of our study.

\subsection{Rules of calculus and covariance of the Langevin equation}
Consider a change of variables $u(t) = U(x(t))$ of the process $x(t)$, where $U$ is a differentiable and invertible function.
Natural questions are: Is it valid to use the chain rule to compute $\dot u(t)$? What is the Langevin equation governing the process~$u(t)$?
In discrete time, defining $u_k\equiv u(t_k) = U(x_k)$, one expresses $\Delta u \equiv u_{k+1} - u_k = U(x_k+\Delta x)-U(x_k)$ using a
Taylor expansion in powers of the increment $\Delta x$, 
\begin{equation}
\Delta u 
= U'(x_k) \Delta x + \frac{1}{2} U''(x_k) \Delta x^2+ {O}(\Delta x^3)
\label{eq:devDeltauStrato}
\end{equation}
that, using $\bar x_k\stackrel{\st}{=} x_k+\frac{1}{2}\Delta x$  and $\Delta x=O(\Delta t^{1/2})$,
becomes
\begin{equation}
\frac{\Delta u}{\Delta t}
\stackrel{\st}{=} 
U'(\bar x_k) \frac{\Delta x}{\Delta t} + {O}(\Delta t^{1/2})
\; . 
\label{eq:chain-rule-Strato}
\end{equation}
In the continuous-time limit, the terms of order $\Delta t^{1/2}$ and 
higher in \eeref{eq:chain-rule-Strato} are negligible, and one recovers the usual chain rule, $\dot u(t) \stackrel{\st}{=} U'(x(t)) \dot x(t)$,
within the Stratonovich scheme.

To determine the evolution equation verified by $u(t)$ in the Stratonovich scheme, one defines
$\bar u_k = (u_k+u_{k+1})/2$.
Inserting~\eref{eq:DeltaxDeltatStrato} into~\eref{eq:chain-rule-Strato}%
\footnote{And using $h(\bar x_k)\stackrel{\textnormal{\tiny{S}}}{=}h( U^{-1}(\bar u_k))+O(\Delta t)$ valid for any function $h$.},
 the time-discrete equation follows
\begin{equation}
\frac{\Delta u}{\Delta t} = F(\bar u_k) + G(\bar u_k) \eta_k + 
{O}(\Delta t^{1/2})
\label{eq:eqLu-discrete}
\end{equation}
where $F(u)$ and $G(u)$ are the force and the noise amplitude of the Langevin equation verified by $u(t)$, defined as 
$F(U(x))=U'(x) f(x)$ and $G(U(x))=U'(x) g(x)$.
% , and evaluated at the $\bar u$ that corresponds to the 
% mid-point $\bar x_k$, that at the relevant order is $\bar u(t_k) = [u(t_k)+u(t_{k+1})]/2$.

Consistently with the chain rule, at leading order in $\Delta t$ and
using the inverse function that leads from $u(t)$ to $x(t)$, one recovers \eeref{eq:DeltaxDeltatStrato} for the original 
process $x(t)$ from \eeref{eq:eqLu-discrete}, thus proving that a Stratonovich-discretized Langevin equation is covariant. 
In short, with the Stratonovich discretization, 
the standard chain rule of differential calculus can be used without 
caution most of the time, even though none of the manipulated objects is actually differentiable!
(These properties are generalized to other {\it linear} discretization schemes $\bar x_k=x_k+\alpha \Delta x$, including the celebrated Itō one, 
once the rules of calculus are modified 
appropriately~\cite{janssen_renormalized_1992,kampen_stochastic_2007,gardiner_handbook_1994}.)

The subleading terms of order $\Delta t^{1/2}$ in Eqs.$\:$\eref{eq:chain-rule-Strato} and~\eref{eq:eqLu-discrete} 
show that the chain rule or the Langevin equation for $u(t)$ are not exact at finite $\Delta t$, but become valid only in the continuous-time limit.
Computing such terms explicitly improves, for instance, the precision of numerical algorithms (inevitably defined in discrete time, see \emph{e.g.}~\cite{milshtejn_approximate_1974,milshtejn_approximate_1975,kloeden_numerical_1995,mannella_integration_2002,jentzen2011taylor,kloeden2012numerical}).
More importantly
for our purposes, we will show that these subleading terms are responsible for the breakdown of 
covariance in the standard path-integral formalism.  This raises a
natural question that we address in the following section: whether there exists a discretization scheme for which the Langevin equations be exactly covariant, that is up to an arbitrary order in $\Delta t$.

\subsection{Improved 
covariant discretization}
A $g$-dependent discretization scheme of the form
\begin{align}
\quad\:\bar x_k& \stackrel{\beta_g}{=}  x_k+\tfrac 12 \Delta x +\betag(x_k) \Delta x^2
\; , 
\label{eq:def-cov-discr-Tg}
\\[1mm]
\betag(x)&=\frac{1}{24}\frac{g''(x)}{g'(x)}-\frac{1}{12}\frac{g'(x)}{g(x)}
\; , 
\label{eq:expr-betag}
\end{align}
yields an evolution equation~\eref{eq:eqLu-discrete} for $u(t)$ valid up to order $\Delta t$, namely one more order in $\Delta t^{1/2}$ than the Stratonovich one. The ensemble of points $\{x_\ell\}$ generated by one such Langevin equation are shown
with diamonds in Fig.~\ref{fig:compar-discr-path}.
Such a scheme, that we call \emph{covariant discretization} (or for short $\beta_g$-discretization), serves as a starting point for our construction of the path integral, where the argument of every function in the action will be understood as discretized according to \eeref{eq:def-cov-discr-Tg}.
As described in Appendix~\ref{sec:exactcovdiscr}, a full series in powers of $\Delta x$ can be added to~\eeref{eq:def-cov-discr-Tg} in order to yield a chain rule~\eref{eq:eqLu-discrete} that is exact to all orders in $\Delta t$ (see the expansions of Eqs.$\:$\eref{eq:Tfg-discretization-def} or~\eref{eq:defdeltaofDeltax}).
Yet, as we show later, the sole additional contribution $\betag(x)\Delta x^2$ in \eeref{eq:def-cov-discr-Tg} is sufficient 
%
% to cure the path integral from its problems upon changing paths.
%
to immunize path integrals against {the problems caused by} nonlinear manipulations. 

When used in the discrete Langevin equation~(\ref{eq:DeltaxDeltatStrato}), the covariant discretization~(\ref{eq:def-cov-discr-Tg})-(\ref{eq:expr-betag}) yields the same equation as the Stratonovich one in the $\Delta t \to 0$ limit: these two schemes are equivalent.
An essential aspect of our construction is that such an equivalence becomes wrong in the path-integral action: as we will show, the covariant and the Stratonovich schemes are not equivalent discretizations when used in the Lagrangian.

Finally, note that for the covariant discretization to be well defined, we assume that the dynamics ensures that $x(t)$ stays in an interval of the real line where  $g(x)>0$  and  $g''(x)/g'(x)$ remains finite.

\section{Probability distribution function of a trajectory} 
\label{sec:probtraj}

We now focus on the construction of the path probability $\mathbb{P}_X[\{x_\ell\}]$.
Such an expression is handy since with it one can directly compute the average of any 
observable of interest,  $\mathcal F[x]$, as the path-integral
$
 \big\langle \mathcal F[x]\big\rangle
  =
  \int \mathcal Dx\,\mathcal F[x] \, \mathbb P_X[x]
  \,
  P_{\text{i}}(x(0)\vphantom{o^1})
$
interpreted in the Feynman sense~\cite{feynman_space-time_1948}: 
a sum over {all} possible trajectories in discrete time
with the measure defined as ${\mathcal D} x = \prod_{k=0}^N dx_k$. 
The initial condition is sampled by~$P_{\text{i}}$. 
We will compare in Section~\ref{sec:comparotherconstrc} our expression for the path probability to the many existing results in the literature. % after its derivation.

% \textcolor{green}{Should we talk about Feynman here?}

% \textcolor{red}{We should check the notation, here and everywhere}

\subsection{Propagator}
%
% Using the notation
% $x_t$  for the variable defined in discrete time $t=0$, $\Delta t$, $2 \Delta t$, $\ldots$ 
%
The path probability of a trajectory is inferred from the infinitesimal propagator 
$
\mathbb P_X(x_1|x_0)
\equiv
\PP(x_1,\Delta t|x_{0},0)
$ 
for the first time step, defined as the conditional probability that $x(\Delta t)=x_1$ at time $t_1=\Delta t$, given $x(0)=x_0$ at time $t_0=0$.
Indeed, the full trajectory p.d.f.~reads 
\begin{eqnarray}
\mathbb{P}_X[\{x_\ell\}] =  
\!\!
\prod_{{0\leq k< N}}
  \!\!\!
 % \dd x_t\,
  \PP(x_{k+1},t_{k+1}|x_{k},t_k)
%\,
%&  \stackrel{\Delta t\to 0}\longrightarrow
%\,
%  \mathcal Dx
%  \,
%\nonumber\\
%
%&\equiv& 
%
\equiv
  {\mathcal N}_{\!X}[\{x_\ell\}]
  \,
\ee^{-S_{\!X}[\{x_\ell\}]}
\; .
\label{eq:Feynman-path-int}
\end{eqnarray}
In the above formula we used a standard representation in which the path probability is written as the 
product of the exponential of an action $S[x]$ and a normalization factor ${\mathcal N}[x]$. Clearly, this 
separation is not unique as factors can be exponentiated in the action or 
{\it vice versa}. We adopt the convenient choice~\cite{deininghaus_nonlinear_1979,hanggi_path_1989}
\begin{align}
  {\mathcal N}_{\!X}[\{x_\ell\}]
  \equiv
  \!\!\!
  \prod_{0\leq k< N\;} 
  \!\!\!
  \frac{1}{\sqrt{4\pi D\Delta t}}
\:
  \frac
  1
  {|g( x_{k+1})|}
\label{eq:Jacobialphabetaendpoint}
\end{align}
of discretizing at the endpoint, that is different from another standard convention in which the prefactor is
discretized at $\bar x_k$~\cite{graham_statistical_1973,wissel_manifolds_1979,lau_state-dependent_2007,itami_universal_2017}. The reason for adopting \eref{eq:Jacobialphabetaendpoint}
instead of the latter
% a product of
% $
% 1/
% |g(\bar x_k)|
% $
is that when changing paths from $\{ x_\ell \}$ to $\{u_\ell\}$, the corresponding Jacobians and the conversion of the 
prefactor bring out factors 
$|{U'(x_{k+1})}|$  and $1/|{U'(x_{k+1})}|$ ($0\leq k< N$)
that cancel one by one, including at time boundaries. 
Another choice would lead to a normalization prefactor that is not covariant, implying that upon a change of variables extra terms coming from the prefactor would impact the action (see for instance~\cite{Cugliandolo-Lecomte17a}).
The choice~(\ref{eq:Jacobialphabetaendpoint}) allows one to focus on the sole transformation properties of the action in the exponential.
The full form~\eref{eq:Feynman-path-int} is inferred from the elementary propagator for the first time step  that we write~as
%
%$\mathbb P_X(x_1|x_0) = \ee^{-\deltaS_X^{\Delta t}}\big / \big[ \sqrt{4\pi D\Delta t} \,|g( x_{1})|\big]$.
%
\begin{align}
%\!\!\!
\mathbb P_{X}(x_1|x_0) 
%\; 
%\stackrel{{\betag}_{\vphantom{I}}} 
=
\;  
&
\frac
{
1 
}
{\sqrt{4\pi D \Delta t}\,|g(x_1)|}
\!
\;
\ee^{-\deltaS_{\!X}^{\Delta t}}
\; . 
\label{eq:PXx0x1_infaction_endpoint-discretization}
\end{align}

\subsection{Stratonovich action}
Following well-known routes \cite{Strato-original_1962,stratonovich1971probability,wissel_manifolds_1979,graham_statistical_1973,dekker_functional_1976,Janssen1976,arnold_symmetric_2000,lau_state-dependent_2007,aron_dynamical_2016,Cugliandolo-Lecomte17a,itami_universal_2017}, one finds that, in the 
Stratonovich scheme, the elementary contribution $\deltaS_{\!X}^{\Delta t}$ to the action for the first time step between $x_0$ and $x_0+\Delta x$ reads
\begin{align}
\!\!
\deltaS_{\!X}^{\Delta t} 
\stackrel{\st}{=}
&
\:
 \frac 12 \frac{\Delta t}{2D} 
\Big[
  \frac
  {\frac{\Delta x}{\Delta t}-f (\bar x_0 )}
  {g(\bar x_0)}
\Big]^2
\!
+
\frac{\Delta t}{2}\Big[ f'(\bar{x}_0)  -\frac{ f(\bar{x}_0) g'(\bar{x}_0)}{ g(\bar{x}_0)} \Big]
\nonumber
\\
&
\qquad\qquad\qquad
+\frac{D}{4}  \big[2 g'(\bar{x}_0){}^2-g(\bar{x}_0) g''(\bar{x}_0)\big]  \Delta t 
\;,
%%%
%%% DO NOT REMOVE: from langevin_non-linear-transformation_Fred-discr_generic-alpha-beta_end-point-prefactor.nb 
%%%
\label{eq:resinfinitesimalpropagx0bS}
\end{align}
that in the continuous-time writing yields the action
%%%%%%which gives the continuous-time writting for the action
%
%
\begin{align}
 S^{\st}_{\!X}[x] 
\stackrel{\st}{=} 
\int_0^{\tf} \!\! \dd t \,
\bigg\{
&
 \frac{1}{4D} 
\Big[
  \frac
  {\dot x-f(x) } {g(x)}
\Big]^2
+
\frac{1}{2} f'(x) -\frac{ f(x) g'(x)}{2 g(x)}
\nonumber
\\
&
\quad\quad\quad\quad\quad
+\frac{D}{4}  \big[2 g'(x)^2-g(x) g''(x)\big]  
\
\bigg\}
\; . 
\label{eq:OMaction_Strato}
%\\%[-7mm]
% \nonumber  %
\end{align}
The reader can easily verify that this continuous-time action
is not covariant.
By this we mean that under a change of variables $x\mapsto U(x)$,
and using the chain rule $\dot u=\dot x\, U'(x)$,
one does not find the correct action $S^{\st}_U$, that has the same form as $S^{\st}_{\!X}$  with the replacements $f \mapsto F$ and $g\mapsto G$ (and similarly if one tries to reconstitute $S^{\st}_{\!X}$ from $S^{\st}_U$).
Such problems were noted in the early developments of path integrals (see {\it e.g.}~\cite{edwards_path_1964,gervais_point_1976,Sa77}).
% and addressed several times afterwards~\cite{deininghaus_nonlinear_1979,LaRoTi79,ApOr96,aron_dynamical_2014_arxiv1,Cugliandolo-Lecomte17a}.
%
The reason, originally identified in~\cite{LaRoTi79}, is actually simple: going for instance from $u$ to $x$, 
the dominant term (of order $\Delta t^0$) in $\deltaS_U^{\Delta t} $ is $\frac{\Delta t}{4D G(\bar u_0)^2} [\frac{\Delta u}{\Delta t}]^2$, see \eeref{eq:resinfinitesimalpropagx0bS}.
Changing variables, one uses the Stratonovich-discretized chain rule~\eref{eq:chain-rule-Strato}. The dominant term in~\eeref{eq:chain-rule-Strato} yields the expected dominant term  $\frac{\Delta t}{4D g(\bar x_0)^2} [\frac{\Delta x}{\Delta t}]^2$  in $\deltaS_{\!X}^{\Delta t} $, but the rest in~\eeref{eq:chain-rule-Strato} yields a double-product contribution $\frac{\Delta t}{2D g(\bar x_0)^2}\frac{\Delta x}{\Delta t}\times O(\Delta t^{1/2})$ which is of order $\Delta t$ and thus \emph{cannot be neglected}.
%
% Instead of showing the failure of the transformation in detail here, we prefer to prove, below, how the covariance is 
% recovered when the covariant discretization is used.
%
The conclusion is simple: in the Stratonovich scheme, using the continuous-time chain rule  $\dot u=\dot x\, U'(x)$ in the action yields a wrong result because the rest $O(\Delta t^{1/2})$ in \eeref{eq:chain-rule-Strato} that could be neglected at the Langevin level (in~\eeref{eq:eqLu-discrete} for instance) cannot be neglected in the action.
While the Stratonovich discretization~\eref{eq:StratoLangevin} was sufficient to render the Langevin 
equation~\eref{eq:DeltaxDeltatStrato} covariant, it fails to play the same role at the path-integral level. 
Changing variables is still possible in~(\ref{eq:OMaction_Strato}) but this involves highly intricate rules (see for instance~\cite{Cugliandolo-Lecomte17a}). 
At this stage, we recall the lesson of Edwards and Gulyaev~\cite{edwards_path_1964}: path integrals are more sensitive to 
discretization issues than Langevin equations, and higher orders in $\Delta t$ than those usually retained, eventually matter. 
This was also noted in \cite{gervais_point_1976,Sa77} in the quantum context, and further discussed in \cite{deininghaus_nonlinear_1979,LaRoTi79,ApOr96,Cugliandolo-Lecomte17a}.

\subsection{A covariant action}
If, instead of writing the infinitesimal action $\deltaS_{\!X}^{\Delta t}$ using the Stratonovich convention as in \eeref{eq:resinfinitesimalpropagx0bS}, one uses the covariant discretization,
\begin{align}
\!\!\!\!
\deltaS_{\!X}^{\Delta t} 
\stackrel{\betag}{=}
&
\:
\frac{\Delta t}{4D} 
\Big[
  \frac
  {\frac{\Delta x}{\Delta t}-f (\bar x_0 )}
  {g(\bar x_0)}
\Big]^2
\!\!\!
+
\frac{\Delta t}{2}\Big[\! f'(\bar{x}_0) -\frac{ f(\bar{x}_0) g'(\bar{x}_0)}{ g(\bar{x}_0)}\Big]
\; ,
\label{eq:resinfinitesimalpropagx0bTg}
%\\%[-7mm] \nonumber  %
%%%
%%% DO NOT REMOVE: from langevin_non-linear-transformation_Fred-discr_generic-alpha-beta_end-point-prefactor.nb 
%%%
\end{align}
where $\stackrel{\beta_g}{=}$ indicates that $\bar x_0$ is $\beta_g$-discretized as in 
Eqs.$\:$\eref{eq:def-cov-discr-Tg}-\eref{eq:expr-betag}.
Compared to the standard Stratonovich scheme ($\betag\equiv0$) one observes that \eeref{eq:resinfinitesimalpropagx0bTg} has less terms: %, thanks to the $\betag$ term. 
the second line in \eeref{eq:resinfinitesimalpropagx0bS} is now absent. 
This means that, in the $\Delta t\to 0$ limit, the covariant and the Stratonovich schemes are not equivalent when writing the action (while they are for the Langevin equation).
This is a signature of the higher sensitivity of the path integral to the details 
of the discretization\footnote{%
Had we kept a term of the form $\gamma(x)\Delta x^3$ in the expansion of \eeref{eq:def-cov-discr-Tg}, this would not have changed the form~\eref{eq:resinfinitesimalpropagx0bTg} 
of $\deltaS_{\!X}^{\Delta t}$ to the order relevant for the path integral (namely, up to $O(\Delta t)$ included). 
The covariant discretization~\eref{eq:def-cov-discr-Tg} thus goes up to the optimal order in powers of $\Delta x$.
}.

The two expressions we have obtained for the infinitesimal action, Eqs.$\:$\eref{eq:resinfinitesimalpropagx0bS} and~\eref{eq:resinfinitesimalpropagx0bTg}, are both valid, \emph{and actually equal}, even though they lead to visually distinct continuous-time writings of the action (compare~(\ref{eq:OMaction_Strato}) with~(\ref{eq:OMactionalphabetatildeTfg}), below).
Changing $\beta_g(x)$ in \eeref{eq:def-cov-discr-Tg} modifies the continuous-time writing of the action (but not that of the Langevin equation).
In the next section we prove that, in contrast to the Stratonovich case, the covariant discretization ensures 
the covariance of the action under a change of path $x(t)\mapsto u(t)=U(x(t))$ through the use of the chain rule.  
%
%We will see that this holds irrespective of whether one follows the correct time-discrete procedure or the naive continuous-time chain rule.

\smallskip
We can draw here a helpful analogy: a multiplicative Langevin process can be described by equivalent but distinct continuous-time writings (depending on the discretization conventions). These are equally valid but only the Stratonovich one benefits from covariance. The same happens for the path-integral: the infinitesimal actions~(\ref{eq:resinfinitesimalpropagx0bS}) and~(\ref{eq:resinfinitesimalpropagx0bTg}) [and their continuous-time writings~(\ref{eq:OMaction_Strato}) and~(\ref{eq:OMactionalphabetatildeTfg})]  are both correct but only~(\ref{eq:resinfinitesimalpropagx0bTg}) benefits from covariance.

\begin{figure}[t]
  \centering\includegraphics[width=.77\columnwidth]{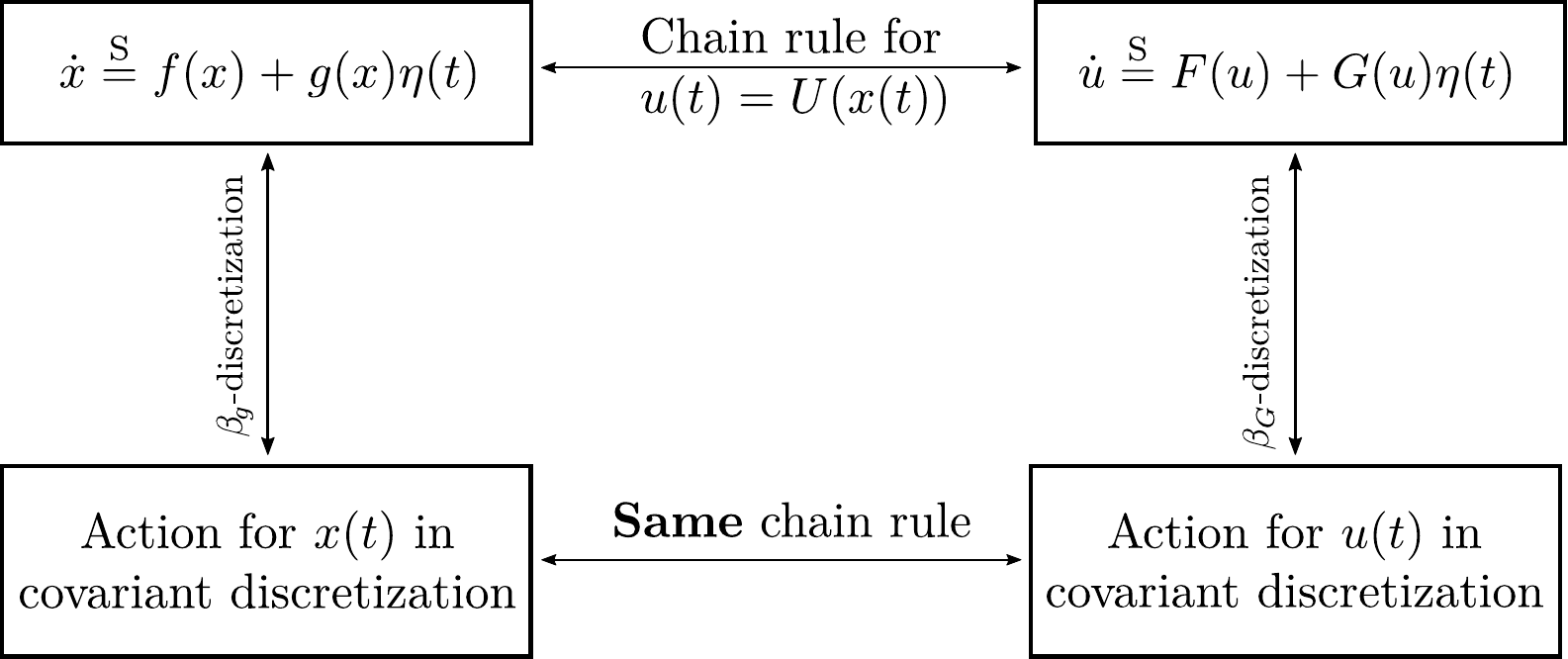}
  \caption{
Schematic representation, for a change of variables $x\mapsto u(x)$, of how the covariant discretization scheme allows one to use the same rules of calculus for a Stratonovich-discretized Langevin equation and for their corresponding covariant Onsager--Machlup and MSRJD actions~\eref{eq:OMactionalphabetatildeTfg} and~\eref{eq:OMactionalphabetatildeMSRJDalt2}.
Such a use of the chain rule would be incorrect in the traditional Stratonovich-discretized actions~\eref{eq:OMaction_Strato} and~\eref{eq:OMactiontildeMSRJD-pure-strato}.
}
\label{fig:diagcomm}
\end{figure}%

\subsection{The proof of covariance}
For convenience we proceed backwards from $u$ to $x$  (see Fig.~\ref{fig:diagcomm}). The infinitesimal propagator for the process $u(t)$ reads
%\footnote{%
%with $G(u)=U'(U^{-1\!}(u))\, G(U^{-1\!}(u))$ where $U^{-1\!}$ is the inverse function of $U$.%
%}
\begin{align}
\!\!\!
\mathbb P_{U}(u_1|u_0) 
\; 
\stackrel{{\betaG}_{\vphantom{I}}} 
=
\;  
&
\tfrac
{
1 
}
{\sqrt{4\pi D \Delta t}\,|G(u_1)|}
\!
\;
\ee^{
\!\!
-\frac 12 \frac{\Delta t}{2D} 
\!
\Big[
  \!
  \frac
  {\frac{\Delta u}{\Delta t}-F (\bar u_0 )}
  {G(\bar u_0)}
  \!
\Big]^2
}
\nonumber\\
&
\qquad
\qquad
\,\times
\ee^{
- \frac 12 \Delta t 
\big[
 \!
 F'\!(\bar u_0) - \frac{F(\bar u_0) G'\!(\bar u_0)}{G(\bar u_0)} 
\big]
}
\; .
\label{eq:resinfinitesimalpropagxdt-fin-U}
\end{align}
We have to show that it yields back the 
 infinitesimal propagator~\eref{eq:PXx0x1_infaction_endpoint-discretization} and action~\eref{eq:resinfinitesimalpropagx0bTg} for the variable $x(t)$ after a generic change of variables. 

First, using that 
$
\mathbb P_X(x_1|x_0) 
= 
|{U'(x_1)}| \; \mathbb P_U(u_1|u_0) 
,
$
% \begin{equation}
% \mathbb P_X(x_1|x_0) 
% = 
% |{u'(x_1)}| \; \mathbb P_U(u_1|u_0) 
% ,
% \label{eq:genericlinkPxPuOM}
% \end{equation}
%(where $\mathbb P_U$ is the propagator for the process $U_t$)
one notices that the  prefactor of the propagator becomes the expected one, \eeref{eq:PXx0x1_infaction_endpoint-discretization}, for the variable 
$x(t)$, thanks {to} the end-point discretized prefactor.
Then, the difficulty is to shift from the $\beta_G$-discretized variable $u(t)$ to the $\beta_g$-discretized variable $x(t)$,
but this only requires a correct expansion at $O(\Delta t)$.
With the recipe presented in the Appendix~\ref{sec:ch-var-discrete}, one compares the following routes:
\begin{itemize}
\item[(\textsc{a})] in \eeref{eq:resinfinitesimalpropagxdt-fin-U}, express $\bar u_0$ as a function of $\bar x_0$ and $\Delta x$; expand in powers of $\Delta x=O(\Delta t^{1/2})$ up to order $O(\Delta t)$; and use thus the substitution rules derived in Ref.~\cite{Cugliandolo-Lecomte17a} (and recalled in Appendix~\ref{sec:substitution-rules}) to handle powers of $\Delta x$ of degree higher than 1; 
\item[(\textsc{b})] naively replace $\frac{\Delta u}{\Delta t}$ in \eeref{eq:resinfinitesimalpropagxdt-fin-U} by $U'(\bar x_0) \frac{\Delta x}{\Delta t}$;
 $F(\bar u_0)$ by $U'(\bar x_0) f(\bar x_0)$; and $G(\bar u_0)$ by $U'(\bar x_0) g(\bar x_0)$.
\end{itemize}
Route (\textsc{b}) is in principle completely faulty because it misses many terms of orders $O(\Delta t^{1/2})$ and $O(\Delta t)$, as discussed in Ref.~\cite{Cugliandolo-Lecomte17a}.
However,
 for the chosen $\beta_g$-discretization of \eeref{eq:def-cov-discr-Tg},
%in $\beta_g$-discretization
it correctly matches the outcome of route (\textsc{a}) --~which happens to be the expected infinitesimal propagator $\mathbb P_X(x_1|x_0)$, given by Eqs.$\:$\eref{eq:PXx0x1_infaction_endpoint-discretization} and~\eref{eq:resinfinitesimalpropagx0bTg}.
For other choices of time discretization, including the Stratonovich one, route (\textsc{b}) does not yield the correct result.
%, which is illustrated by the failure of the manipulations of our toy example.

Since taking route (\textsc{b}) amounts to using the standard rules of calculus in the action, we have thus shown that, for the $\beta_g$-discretization~\eref{eq:def-cov-discr-Tg}, the correct rules of calculus in the infinitesimal propagator at small but finite $\Delta t$ become identical to the standard rules of calculus in the 
continuous-time action when taking the $\Delta t\to 0$  limit.
Showing the validity of the chain rule in this limit is simple for differentiable functions and significantly more intricate in a Langevin equation (where discretization issues matter), and it has demanded 
an even higher degree of caution inside the  action, through the use of the covariant discretization~\eref{eq:def-cov-discr-Tg} (see Table~\ref{tab:discretization-vs-situation}).

\begin{table}[]
\centering
\begin{tabular}{|l|l|}
\hline
\textbf{Situation}            & $\!\!$\textbf{\begin{tabular}[c]{@{}l@{}}Required\\ discretization\end{tabular}}                  \\ \hline
$x(t)$ is differentiable      & Any can work                                                                                    \\ \hline
$x(t)$ is a Langevin process, \eeref{eq:eqLangevin}  & Stratonovich, \eeref{eq:StratoLangevin}$\!\!$                              \\ \hline
$x(t)$ is a path in the covariant action, \eeref{eq:OMactionalphabetatildeTfg} or~\eref{eq:OMactionalphabetatildeMSRJDalt2}$\!\!\!$ & Covariant, \eeref{eq:def-cov-discr-Tg}         \\ \hline
$x(t)$ is a path in the standard action, \eeref{eq:OMactionalpha12beta0} or~\eref{eq:OMactiontildeMSRJD-pure-strato}& None works                                             \\ \hline
\end{tabular}
\caption{Minimal discretizations required  
for the chain rule of standard calculus to hold upon a change of variables $u(t)=U(x(t))$.}
\label{tab:discretization-vs-situation}
\end{table}

\subsection{Summary and continuous-time writing}
\label{sec:our-result}
Our main result is the direct construction of the probability of a time-discretized path. It takes the form of a path-integral probability~\eref{eq:Feynman-path-int} with an endpoint-discretized %normalization 
prefactor $\mathcal N[x]$ (\eeref{eq:Jacobialphabetaendpoint}) and a $\beta_g$-discretized time-discrete action read from \eeref{eq:resinfinitesimalpropagx0bTg},
\begin{align}
S_{\!X}^{\Delta t}[x]
\;
\stackrel{\betag}{=}
\!\!\!
\sum_{0\leq k<N}
\!\!%\!
\bigg\{
\!\!
&
\:
 \tfrac 12 \tfrac{\Delta t}{2D} 
\Big[
  \tfrac
  {\tfrac{\Delta x}{\Delta t}-f (\bar x_k )}
  {g(\bar x_k)}
\Big]^2
+
\tfrac{\Delta t}{2}\Big[ f'(\bar{x}_k) -\tfrac{ f(\bar{x}_k) g'(\bar{x}_k)}{ g(\bar{x}_k)}\Big]
\bigg\}
.
\label{eq:Tgaction}
%\nonumber 
%\\[-7.5mm] \nonumber  %
%%%
%%% DO NOT REMOVE: from langevin_non-linear-transformation_Fred-discr_generic-alpha-beta_end-point-prefactor.nb 
%%%
\end{align}
Taking the continuous-time limit, 
the path probability $\mathcal N[x] \ee^{-S[x]}$ of a trajectory $[x(t)]_{0\leq t\leq \tf}$ that evolves according to the Langevin equation~\eref{eq:eqLangevin} (understood in the Stratonovich sense)
% (that is equivalent to the $\beta_g$ one at the level of Langevin equation in the continuous-time limit) 
has an action given by 
\begin{equation}
\label{eq:OMactionalphabetatildeTfg}
  S_{X\!}[x]
\;
  \stackrel{\betag}{=}
  \int_0^{\tf}
  \!\!
  \dd t\,
  \bigg\{
   \frac{1}{4D} 
  \bigg[\frac
  {
  \dot x-f (x) 
  }
  {g(x)}\bigg]^2
  +
  \frac 12  f'(x)
  -
  \frac 12
  \frac{f(x) g'(x)}{g(x)}  
\bigg\}
\end{equation}
which is a short-hand writing for the discrete expression~(\ref{eq:Tgaction}).
Such continuous-time writing of the action turns out to coincide with the result of Refs.~\cite{Strato-original_1962,horsthemke_onsager-machlup_1975,graham_path_1977}.
In our formulation, it benefits from an essential feature: it is \emph{covariant} under the change \emph{of non-differentiable paths}, in the sense that the path probability of a process $u(t)=U(x(t))$ has a $\beta_G$-discretized action $S_{U\!}[u]$ that is inferred from the action \eeref{eq:OMactionalphabetatildeTfg} for $x(t)$ by merely passing from the variable $x$ to $u$ through the use of the standard chain rule of calculus, see Fig.~\ref{fig:diagcomm}. 
Such property is verified in the continuous-time writing of \eeref{eq:OMactionalphabetatildeTfg} (in a computation that is valid for differentiable paths); but its actual proof is done in discrete time, as presented in the previous paragraph, because the action we are interested in describes the path probability of non-differentiable trajectories, through \eeref{eq:Feynman-path-int}.

Besides, if we were to read the action~(\ref{eq:OMactionalphabetatildeTfg}) as Stratonovich-discretized, the resulting expression would be incorrect: 
as directly checked, the summand of \eeref{eq:Tgaction} evaluated for a Stratonovich-discretized $\bar x_k$ and a covariant-discretized $\bar x_k$ differ by non-constant terms of order $\Delta t$, that cannot be discarded in the $\Delta t \to 0$ limit.
This explains why continuous-time derivations of the action~(\ref{eq:OMactionalphabetatildeTfg}), such as the one of Graham~\cite{graham_path_1977}, are not amenable to an easy reconstruction of the path probability. For instance, in a subsequent work, Graham and Deininghaus~\cite{deininghaus_nonlinear_1979} succeeded to do so, but at the price of multiplying the trajectory weight with a correction prefactor that is tuned in order to ensure covariance and probability conservation. In contrast, the construction we bring forward is self-contained and establishes that the covariant action~(\ref{eq:OMactionalphabetatildeTfg}) simply has to be read with the covariant discretization scheme.

\section{Comparison to other path-integral constructions}
\label{sec:comparotherconstrc}

\subsection{Different approaches}
To write the explicit path probability of a trajectory, the time-slicing procedure can be implemented in a variety of ways and, 
within the realm of stochastic processes, this  was carried out 
in~\cite{onsager_fluctuations_1953,graham_statistical_1973,Janssen1976,dominicis_techniques_1976,dekker_functional_1976,arnold_symmetric_2000}. The constructions proposed by these authors are not fully satisfactory 
as they all suffer from the same problem: their actions are neither covariant at the discretized nor at the continuous-time levels.
A classical choice in these papers consists in writing the action in Stratonovich and
discretizing the normalization prefactor $\mathcal N[x]$ in $\bar x_k$  (replacing $g(x_{k+1})$ by $g(\bar x_k)$ in~\eref{eq:Jacobialphabetaendpoint}). This leads to the continuous-time writing
\begin{align}
\label{eq:OMactionalpha12beta0}
\!\!\!\!
  S[x(t)]
  \stackrel{\st} = 
\frac 12
\!
  \int_0^{\tf} 
\!\!
 \dd t 
\,
  \bigg\{
   \frac{1}{2D} 
  \bigg[\frac
  {\dot x-f (x)+ D \,g(x)g'\!(x)}
  {g(x)}\bigg]^2
  +
  f'(x)
\bigg\}
\end{align}
which is not covariant under the standard rules of calculus. Several authors tried to 
cure this problem. We summarize these attempts, and why we think that the 
goal was not fully achieved, in the next paragraph.
%
%%%The lesson to draw is actually simple: either one sticks to stochastic calculus and 
%%%forgets about path integrals that cannot accommodate  nonlinear changes of integration fields, or %%%one attempts to cure path integrals. 
%
R.~Fox also put forward a path-integral construction that relies on considering a colored instead of a white noise, with a finite correlation time $\tau$~\cite{fox_functional-calculus_1986,fox_stochastic_1987}. Although this approach has the advantage to handle more regular paths, the $\tau\to 0$ limit yields back a Stratonovich-discretized action and is not covariant.

\subsection{Covariant approaches}
The first important progress in solving this problem is due to Stratonovich~\cite{Strato-original_1962,stratonovich1971probability}, who constructed a covariant continuous-time action, whose writing is the same as \eeref{eq:OMactionalphabetatildeTfg}. 
Horsthemke and Bach~\cite{horsthemke_onsager-machlup_1975} and Graham~\cite{graham_path_1977} independently derived the same action in one dimension, and Graham further achieved the same program in dimension larger  than one.
What they did was to build a path integral with an action expressed in continuous time that is consistent with the underlying Langevin equations, and that can be blindly manipulated with the usual rules of differential calculus, as if the paths were differentiable.  
However their construction of the path integral is built from locally optimal differentiable paths. The action~\eref{eq:OMactionalphabetatildeTfg} thus bears different meanings in the mentioned references and in this work.
Related works in mathematics have made such an approach rigorous.
Either using changes of path probability~\cite{tisza_fluctuations_1957,durr_onsager-machlup_1978} or more direct techniques (see the work of Takahashi and collaborators~\cite{takahashi_probability_1981,hara_lagrangian_1996} and of Capitaine~\cite{capitaine_onsager-machlup_2000}), the idea is to determine the most probable path\footnote{A ``path'' seen as an infinitesimal tube around a \emph{differentiable} trajectory.} going from one point to an other as extremizing an Onsager--Machlup covariant action. Such constructions are possible but do not provide the path probability of an arbitrary non-differentiable trajectories (which is the aim of our theoretical physicist's construction).

In the immediate aftermath of Graham's result, the search for an ambiguity-free definition of the path probability of a trajectory began. This is commented by Graham in \cite{graham_covariant_1977}, and it triggered more works \cite{weiss_operator_1978,kerler_definition_1978,LaRoTi79,langouche_comment_1980,Tirapegui82,deininghaus_nonlinear_1979, AlDa90,ApOr96,kleinert_path_2009,aron_dynamical_2014_arxiv1,Cugliandolo-Lecomte17a} in the direction of finding a proper discretization, the continuum limit of which would fall back on the action~\eref{eq:OMactionalphabetatildeTfg}. This problem  was not solved until this paper:
we have found for this action an explicitly discretized picture that plays an analogous role to the Stratonovich rule (\eeref{eq:StratoLangevin}) for Langevin equations.
%
%and that actually endows the path-integral weight with the meaning of a non-ambiguous time-discrete path probability.
%
In the context of quantum mechanics in curved spaces, DeWitt~\cite{dewitt_dynamical_1957} followed a construction where the action is evaluated along a succession of infinitesimal optimal trajectories that obey Euler--Lagrange equation
--~a construction that also has been made rigorous by mathematicians~\cite{inoue_integral_1985,andersson_finite_1999}.
Such a procedure yields the same visual limit as the action~\eref{eq:OMactionalphabetatildeTfg} but endows it with  a completely different meaning.

The covariant discretization that we propose in fact provides a step towards extending stochastic calculus to path integrals, by defining the time integral of the action~\eref{eq:OMactionalphabetatildeTfg} through a procedure generalizing the usual stochastic integral.
From our physicist's viewpoint, stochastic calculus provides a definition of the integral 
   $\int \dd t\, \big[ A(x) + B(x) \dot x]$
in a limiting procedure that involves a careful choice of discretization, together with being compatible with continuous-time rules of differential calculus (the standard chain rule). 
The construction we put forward allows one to do the same for  
   $\int \dd t\, \big[ A(x) + B(x) \dot x + C(x) \dot x^2]$
inside an exponential.

\section{Martin--Siggia--Rose--Janssen--de$\,\,$Dominicis (MSRJD) path-integral formulation}
\label{sec:msr}

Since the early formulation of quantum mechanics in terms of path integrals, there have been two equivalent 
{expressions} for {the} transition amplitudes. One, that we have just discussed extensively, involves a single position field. An alternative 
one involves an additional conjugate momentum field. The latter can be removed or included at will by Gaussian integration. A mirror image of the 
auxiliary momentum field exists for stochastic dynamics: the alternative to the original Onsager--Machlup formulation 
{is the MSRJD approach~\cite{Martin1973,kubo_fluctuation_1973,dominicis_techniques_1976,Janssen1976,DeDominicis1978}} and 
involves an additional so-called  response field. The purpose of this section is to extend our findings to this 
{formalism}. 
Again, we adopt the language of stochastic dynamics, but our results equally apply to quantum mechanics.

\subsection{Continuous-time MSRJD covariant action}
\label{sec:response-fields-as}
In the MSRJD approach 
one introduces a response field $\hat x(t)$  to represent the trajectory weight in a manner 
that allows one, for instance, to get rid of some non-linearities of the action~\eref{eq:OMactionalphabetatildeTfg}.
Physics-wise, {this setting facilitates} 
the study of correlations and response functions on an equal footing, and to linearize (to some extent) possible symmetries of the process 
under scrutiny (time-reversal, rapidity reversal, {\it etc}).
We now present our result for the covariant MSRJD action before describing its construction and its full time-discrete implementation.

In the covariant discretization scheme of \eeref{eq:def-cov-discr-Tg}, the action
\begin{align}
  S[\hat x,x]
  \stackrel{\betag}=
  \!
  \int_0^{\tf}\!\! \dd t\,
   & 
   \Big\{
  \hat x
  \big(
  \dot x-f (x) + D \,g(x)g'(x)
  \big)
   -
  D g(x)^2 \hat x^2
  \nonumber\\
&
\;\:
  +
  \frac 12  f'(x)
  -
  \frac D4
   g'(x)^2
  -
  \frac 12
   \frac{g'(x)}{g(x)}\,\dot x
\Big\}
\label{eq:OMactionalphabetatildeMSRJDalt2}
\end{align}
describes the path probability measure as
$
  \mathcal Dx
  \,  
  \mathcal D\hat x
  \:
  \ee^{-S[\hat x, x]}
 $.
In this path integral one can directly change variables covariantly using the standard chain rule and avoiding any Jacobian contribution.
In continuous time, this property is tediously checked by direct computation using the chain rule of calculus together with the correspondence $\hat x(t)=U'(x(t))\, \hat u(t)$ 
between response fields.
In contrast, the historically derived MSRJD action in Stratonovich discretization  reads
\begin{align}
\label{eq:OMactiontildeMSRJD-pure-strato}
  \!\!
  \int_0^{\tf}\!\!\! \dd t \
  \Big\{
&
  \hat x
  \big(
  {
  \dot x-f (x) + D \,g(x)g'\!(x)
  }
  \big)
  \! - \! 
  D g(x)^2 \hat x^2
  \! + \! 
   \frac 12  f'\!(x)
\!
\Big\}
\end{align}
and applying the chain rule to it leads to inconsistencies~\cite{aron_dynamical_2016}.

\subsection{Discretized MSRJD action}
\label{sec:discretized-msrjd}
To construct the MSRJD representation, one rewrites the infinitesimal 
propagator for $x(t)$ (Eqs.$\:$\eref{eq:PXx0x1_infaction_endpoint-discretization} and~\eref{eq:resinfinitesimalpropagx0bTg}) by using at every time step a Hubbard--Stratonovich transformation of the form
$
  \sqrt{2\pi/a}
\:
  \ee^{-\frac 12 \frac{b^2}{a}}
=
  \int_{i \mathbb R}
  d\hat x
  \:
  \ee^{\frac 12 a \hat x^2 -  b \hat x}
$
for the following choice of parameters $a$ and $b$
\begin{align}
  a
&
  =
  {2 D g(\bar x_t)^2}
\,
  {\Delta t}
  \; , \quad
\label{eq:HStransfoa}
  b
  =
  \Big[  {\frac{\Delta x}{\Delta t}-f (\bar x_t )}
  \Big]\,\Delta t
  \; ,
\end{align}
which gives
\begin{align}
 \mathbb P(x_1|x_0) 
  &
    \stackrel{\betag} 
    = 
\:
\big| \tfrac{g(\bar x_0)}{g(x_1)} \big|
  \int_{i\mathbb R}
\!\!
  d\hat x_0
\:
\ee^{-\deltaS[\hat x_0,\bar x_0]}
\; , 
\label{eq:MSR_infinit-propag_endpoint}
%\end{align}
%\begin{align}
%
\\[-.2mm]
\deltaS[\hat x_0,\bar x_0]
&
\stackrel{\betag} 
=
\Delta t
\, 
\Big\{
\hat x_0
\big[
 \tfrac{\Delta x}{\Delta t}-f (\bar x_0 )
\big]
-
  D g(\bar x_0)^2 \hat x_0^2
\nonumber
\\[-1.5mm]
&
\qquad\qquad\qquad\
  +
  \frac 12  f'\!(\bar x_0)
  -
  \frac 12
  \frac{ g'\!(\bar x_0)}{g(\bar x_0)}  f(\bar x_0)
\Big\}
\; , 
\label{eq:infinitesimalactionMSRJD}
\end{align}
which completely encodes the continuous-time expression%
\footnote{%
Note from \eeref{eq:MSR_infinit-propag_endpoint} the appearance, in the discretized expression for the probability of a path, of a normalization prefactor
$\NN_{\mkern-1.5mu\textnormal{\tiny{MSR}}}[x(t)]=\prod_{0\leq k< N\!}
  \big| \frac{g(\bar x_k)}{g(x_{k+1})} \big|$
 in front of the exponential weight. This $\NN_{\mkern-1.5mu\textnormal{\tiny{MSR}}}$ warrants that a change of path in the action~\eref{eq:OMactionalphabetatildeMSRJDalt2} induces no spurious contribution coming from the Jacobian $|{U'(x_1)}|$ in
$
\mathbb P_X(x_1|x_0) 
= 
|{U'(x_1)}| \; \mathbb P_U(u_1|u_0) 
$. 
}%
\begin{align}
\label{eq:OMactionalphabetatildeMSRJDalt}
  \tilde S[\hat x,x]
  \stackrel{\betag}
  =
  \int_0^{\tf}\!\! \dd t\,
  \Big\{
  \hat x\,
  \big(
  \dot x & -f (x) 
    \big)
  -
  D g(x)^2 \hat x^2
\nonumber
\\[-2.5mm]
&
  +
  \frac 12  f'(x)
  -
  \frac 12
  \frac{ g'(x)}{g(x)}  f(x)
\Big\}
\;.
\end{align}
Up to a translation of the field $\hat x(t)$ by $-g'/(2g)$, one 
 recovers \eeref{eq:OMactionalphabetatildeMSRJDalt2}.
The symbol $\betag$ over the equality sign means that functions of the variable $x$ are $\beta_g$-discretized, \emph{i.e.}~evaluated at~$\bar x_k$.
The field $\hat x(t)$ is not discretized in the same way as the field $x(t)$ is: a variable $\hat x_t$ is introduced at each~$t$ and merely associated to $\bar x_k$.
The proof of the covariance presents more intricate issues than for the Onsager--Machlup action, and is sketched in Appendix~\ref{sec:MSRJD-deriv-covar}.

\section{Summary and outlook}
\label{sec:summary}

When dealing with fluctuating signals as encountered in quantum mechanics or stochastic processes, whose shared trait is non-differentiability, physicists rely on a triptych of methods: solving a 
linear problem involving an operator (Schrödinger or Fokker--Planck equations), resorting to stochastic calculus (Langevin equations), or {using} 
path integrals (field theory). As we have discussed, there is a vast number of operations for which path integrals have been known to be badly flawed. This surely explains why path integrals never became a tool of choice for mathematicians working on similar problems. What we have shown in the present work is how to construct a path-integral calculus that directly manipulates physical paths and that is devoid of what we view as its biggest flaw. 
%
%It is now possible to manipulate well-defined path integrals with nonlinear changes of fields {making no} errors.
%
 It is our belief that our proposed construction should not only trigger a revival of interest on the mathematics side, but also {on the physics one}. 
Mathematics-wise, though we would not blush with embarrassment about our physicist's derivation, it is almost certain that many more steps are needed to bring our building of covariant path integrals on a rigorous par with other aspects of stochastic calculus. Physics-wise, we see immediate consequences, and open questions. Among the 
{former}, given the pedagogical importance of path integrals in higher education, we would advocate strongly in favor of our presentation (which time and efforts will surely smoothen and hopefully simplify) rather than in existing ones which suffer from well-known problems.  Second, given the lack of control, so far, in nonlinear manipulations of fields, which have been put to work in so many areas, it seems like a necessity to return to these and sort out whether and how path-integral based results are altered by taking our corrected formalism into account. 
Transformations of the action based on the chain rule, as simple as integrations by parts for instance, are in principle forbidden unless one uses the covariant discretization.
This is especially important in areas of physics where no alternative to path integrals exist (like in path-integral based quantization issues). This brings us to future research directions, which we briefly list: What about higher space dimensions?, What about supersymmetries?, What about field theories expressed in second quantized form with coherent-states fields?

\medskip

\emph{Acknowledgements.}
VL~acknowledges financial support from the ERC Starting Grant No. 680275 MALIG, the UGA IRS PHEMIN project, the ANR-15-CE40-0020-03 Grant LSD and the ANR-18-CE30-0028-01 Grant LABS.
{LFC is a member of Institut Universitaire de France.}
The authors thank H.~J.~Hilhorst and H.~K.~Janssen for very useful discussions.

%\appendix
\section*{Appendices}
\addcontentsline{toc}{section}{Appendices}

\renewcommand{\thesection}{A}

%\small

\subsection{An exact covariant discretization of the Langevin equation}
\label{sec:exactcovdiscr}
Since the path-integral formulation requires higher orders in $\Delta t$ 
 than usually, it appears crucial to find a discretization scheme {{that is}} consistent with the chain rule to a high-enough order. Fortunately, such a scheme can be found, and this is one of the main results in this paper. The inspiration comes from the field of calculus with Poisson point processes~\cite{paola_stochastic_1992,di_paola_stochastic_1993,di_paola_ito_1993,kanazawa_stochastic_2012}, though our solution departs from anything  that has already been proposed. We postulate that \eeref{eq:eqLangevin} is to be understood in the form
\begin{equation}\label{eq:NGLangevin}
\Delta x=\Tfg f(x(t))\Delta t+\Tfg  g(x(t))\Delta \eta
\end{equation}
with $\Delta \eta=\eta \Delta t$ and where the operator $\Tfg$ acts on an arbitrary function $h$ as
\begin{equation}
\Tfg h(x)=
\frac{\ee^{\mathcal D(x)\,  \frac{\dd}{\dd x}}-\mathbf 1}{\mathcal D(x) \,  \frac{\dd}{\dd x}} h(x)
=\sum_{n\geq 0}\frac{{\left(\mathcal D(x)\frac{\dd}{\dd x}\right)\!}^n}{(n+1)!} h(x)
\:.
\label{eq:Tfg-discretization-def}
\end{equation}
Here\footnote{%
In the study of Poisson point-processes with multiplicative noise, the appropriate discretization restricts to $\mathcal D(x) = g(x)\Delta \eta$, but in our context the supplemental term $f(x)\Delta t$ is needed.
}
$\mathcal D(x) = f(x) \Delta t + g(x)\Delta \eta\,$ acts as an operator, and it does not commute with $\frac{\dd}{\dd x}$.
When acting on $f$ the operator $\Tfg$ leaves us with a complicated function of both $x(t)$ and $\Delta\eta$, which, in an implicit fashion through \eeref{eq:NGLangevin}, is then a function of $x(t)$ and $\Delta x=x(t+\Delta t)-x(t)$. As is perhaps less obvious than in previous discretization schemes, the $\Delta t\to 0$ limit also gets us back to \eeref{eq:eqLangevin}. This is because $\Delta\eta$, which is of order $\Delta t^{1/2}$, also goes to~$0$.
We remark here that truncating the sum at $n=2$ in~(\ref{eq:Tfg-discretization-def}) one recovers an expression that is close to the Milstein~\cite{milshtejn_approximate_1974,milshtejn_approximate_1975} scheme used in numerical simulations of Langevin equation (one has to discard a term $\propto \Delta t \, f(x)\, f'(x) $ and switch from Stratonovich to Itō calculus).

 The complex appearance of this discretization rule~(\ref{eq:NGLangevin})-(\ref{eq:Tfg-discretization-def}) should not conceal its central property: it is consistent with the chain rule {\it for any finite} $\Delta t$. In other words, when the evolution of $x$ is understood with \eeref{eq:NGLangevin}, one can manipulate a function $u(t)=U(x(t))$ as if it were differentiable, and $\dot u = \frac{\dd u}{\dd t}=\dot x \, U'(x)$  holds in the sense that
\begin{equation}
\frac{u(t+\Delta t)-u(t)}{\Delta t}=\TFG F(u(t))+ \TFG G(u(t)) \Delta \eta
\; , 
\label{eq:CRu}
\end{equation}
where $F(u)$ and $G(u)$ are the force and the noise amplitude of the Langevin equation verified by $u(t)$, defined as $F(U(x))=U'(x) f(x)$ and $G(U(x))=U'(x) g(x)$.

The unpleasant feature of the discretization rule in \eeref{eq:Tfg-discretization-def} is that it is expressed in terms of $\Delta\eta$ rather than in terms of $\Delta x$, as we did in \eeref{eq:StratoLangevin}. This means that  \eeref{eq:Tfg-discretization-def} cannot be used as such in the definition of the path integral in which the noise $\eta(t)$ is eliminated in favor of $x(t)$. We would rather express \eeref{eq:NGLangevin} in terms of a function $\delta(\Delta x)$ such that
\begin{equation}
\Tfg h(x)=h(x+\delta(\Delta x))
\; . 
\end{equation}
An expansion of $\delta$ in powers of $\Delta x$ can be found:
\begin{equation}
\delta(\Delta x)
=
\alpha \Delta x
+\beta(x)\Delta x^2 
+\ldots
\label{eq:defdeltaofDeltax}
\end{equation}
where $\alpha=\frac 12$, $\beta=\beta_g=\frac{1}{24}\frac{g''}{g'}-\frac{1}{12}\frac{g'}{g}$, {\it etc}. We shall henceforth keep the functional dependence of these functions on $g$ explicit.
Keeping in mind that $\Delta x=O(\Delta t^{1/2})$ as $\Delta t\to 0$, 
at minimal order $\delta(\Delta x)=\frac 12 \Delta x$ and we recover the Stratonovich discretization~\eref{eq:StratoLangevin}, for which
the chain rule in \eeref{eq:CRu} is valid with up to an error of order $\Delta t^{1/2}$. Including the $\beta$ term in \eeref{eq:defdeltaofDeltax} with $\beta=\betag$ renders the error of order $\Delta t$
(and so on and so forth when increasing the order of the expansion). Terms of order higher than $\beta$ in~\eref{eq:defdeltaofDeltax} will prove useless for our purpose. 
This is the discretization scheme that we adopted in Eqs.$\:$\eref{eq:def-cov-discr-Tg}-\eref{eq:expr-betag} in the time-slicing procedure involved in constructing our formulation of the path integral.

\subsection{Changing variables while respecting the discretization}
\label{sec:ch-var-discrete}
We explain here the methodology used to manipulate the infinitesimal propagator in the small $\Delta t$ limit, following Ref.~\cite{Cugliandolo-Lecomte17a}.
When passing from one infinitesimal propagator to another,  one needs to reconstitute the $\beta_g$-discretization of the variable $x(t)$ in $\mathbb P_{X\!}(x_1|x_0)$ 
(Eqs.$\:$\eref{eq:PXx0x1_infaction_endpoint-discretization} and~\eref{eq:resinfinitesimalpropagx0bTg})
from the $\beta_G$-discretization of the variable $u(t)$ in  $\mathbb P_{U\!}(u_1|u_0)$ (\eeref{eq:resinfinitesimalpropagxdt-fin-U}).
The idea is to express the time-discrete values $u_0=U(x_0)$, $u_1=U(x_1)$ and $\bar u_0$ appearing in the r.h.s. of \eeref{eq:resinfinitesimalpropagxdt-fin-U} as a function of $\bar x_0$ and $\Delta x$, using
\begin{align}
  \bar u_0 &= U(x_0) + \tfrac 12 [U(x_1 )-U(x_{0})] + \betaG(U(x_0))\,[U(x_1 )-U(x_{0})]^2
  \; , 
  \nonumber
  \\
  x_0 &= \bar x_0 - \tfrac 12 \Delta x - \betag(\bar x_0)\, \Delta x^2
  \; , 
\label{eq:X0}
  \\
  x_1  &= \bar x_0 + \frac 12 \Delta x - \betag(\bar x_0)\, \Delta x^2
  \; .
\label{eq:Xdt}
\end{align}
The strategy is the following: first, use these expressions in \eeref{eq:resinfinitesimalpropagxdt-fin-U}; second, 
expand this equation in powers of $\Delta t$ and $\Delta x$, keeping in mind that the latter is $O(\Delta t^{1/2})$.
The result takes the form
\begin{equation}
\tfrac
{\NN}
{|g(x_1)|}
\!
\;
\ee^{
\!
-\frac 12 \frac{\Delta x^2}{2D g(\bar x_0)^2 \Delta t} 
}
\times
\big[
1+\textnormal{polynomial in $\Delta x$ and $\Delta t$}\:
\big]
\:
\label{eq:expansionpolynomialPxfromPU}
\end{equation}
The fraction in the exponential is $O(\Delta t^0)$ and cannot be expanded; in fact, it defines the dominant order $O(\Delta t^{1/2})$ of $\Delta x$.
The polynomial contains terms of the form $\Delta t^n \Delta x^m$ which are of order $O(\Delta t^{1/2})$ and $O(\Delta t)$.
Higher-order terms ($O(\Delta t^{3/2})$ and higher) can be discarded because they do not contribute to the action.
Many of the terms in the polynomial do not present an obvious $\Delta t\to 0$ limit (\emph{e.g.}~$\Delta x^4 \,\Delta t^{-1}$) but the 
substitution rules derived in~\cite{Cugliandolo-Lecomte17a} allow one to take the continuous-time limit. For completeness,
these are recalled (and slightly reformulated) in Appendix~\ref{sec:substitution-rules}.
The last stage of the procedure consists in reexponentiating the resulting factor $[1+\ldots]$ obtained from \eeref{eq:expansionpolynomialPxfromPU}. 
One then recovers the expected propagator $\mathbb P_{X\!}(\xdt|x_0)$ of Eqs.$\:$\eref{eq:PXx0x1_infaction_endpoint-discretization} and~\eref{eq:resinfinitesimalpropagx0bTg} as announced. 

The same procedure allows one to change variables in the historical action~\eref{eq:OMactionalpha12beta0} but this involves rules of calculus sharing little kindred with the 
standard ones (see~Ref.~\cite{Cugliandolo-Lecomte17a}). The covariant discretization, instead, 
yields back the usual chain rule as $\Delta t\to 0$.

\subsection{Sketch of the derivation of the covariance of the MSRJD path-integral}
\label{sec:MSRJD-deriv-covar}
The actual derivation of the covariance property involves a careful handling of the time-discrete infinitesimal propagator, by analyzing the contributions that arise order by order in powers of $\Delta t$ upon the change of variables $U(t)=u(x(t))$.

One proves that only for the covariant discretization it is valid to naively change variables in Eqs.$\:$\eref{eq:MSR_infinit-propag_endpoint}-\eref{eq:infinitesimalactionMSRJD}: namely, going from the fields $(\hat u, u)$ to $(\hat x, x)$, one can
replace $\frac{\Delta u}{\Delta t}$  by $U'(\bar x_0) \frac{\Delta x}{\Delta t}$,
 $F(\bar u_0)$ by $U'(\bar x_0) f(\bar x_0)$, and $G(\bar u_0)$ by $U'(\bar x_0) g(\bar x_0)$.
Such operations, combined with $\hat u_0=\hat x_0/U'(\bar x_0)$, would normally yield an incorrect result by missing essential contributions of 
order $O(\Delta t^{1/2})$ and $O(\Delta t)$.
{Satisfactorily,} these manipulations {are} correct for our chosen covariant discretization.
The proof follows a procedure similar to the one we presented for the Onsager--Machlup case by comparing a correct route (\textsc{a}) with a naive route (\textsc{b}), with three important caveats:
(\emph{i}) One has to pay attention to the fact that $\hat x_t\sim \Delta t^{-1/2}$ at every time step, as inferred from the scaling of $a$ in the Hubbard--Stratonovich 
transform~\eref{eq:HStransfoa}, implying that the expansions in powers of $O(\Delta t)$ bring in terms that one can be tempted to throw away at first sight;
(\emph{ii}) One has to design additional substitution rules in order to handle powers of $\hat x_0$ larger than~$1$. This is done following a procedure similar to the one of 
Ref.~\cite{Cugliandolo-Lecomte17a} (see Appendix~\ref{sec:substitution-rules});
(\emph{iii}) Unexpectedly, in contrast  to the Onsager--Machlup case exposed previously, the prefactor 
$\big| \tfrac{g(\bar x_0)}{g(x_1)} \big|$ in~\eref{eq:MSR_infinit-propag_endpoint} \emph{brings a Jacobian contribution into the action} upon the time-discrete change of variables 
of route~(\textsc{a}), which compensates precisely a term that is missing when naively substituting $\frac{\Delta u}{\Delta t}$  by $U'(\bar x_0) \frac{\Delta x}{\Delta t}$ along route~(\textsc{b}).

To summarize, we have shown that changing variables in the MSRJD action~\eref{eq:OMactionalphabetatildeMSRJDalt2} can be done following the standard rules of differential calculus, provided that the discrete-time construction of the path-integral weight is performed according to the covariant discretization of Eqs.$\:$\eref{eq:def-cov-discr-Tg}-\eref{eq:expr-betag}
 --~leading to a modified action as compared to the historical Stratonovich-discretized~one.

\subsection{Substitution rules}\
\label{sec:substitution-rules}
Denoting
$
  \lceil{\Delta x^2}\rfloor= 2Dg(\bar x_0)^2 \Delta t
$,
the  substitution rules deduced in~\cite{Cugliandolo-Lecomte17a} can be reformulated as follows
\begin{align}
  \label{eq:Ito2b}
  \Delta x^2 &\mapsto \lceil{\Delta x^2}\rfloor
  \; , 
  \\
  \label{eq:Ito3}
  \Delta x^3 \,\Delta t^{-1} &\mapsto 3 \ \Delta x \ \lceil{\Delta x^2}\rfloor \,\Delta t^{-1}
 \; ,  \\
  \label{eq:Ito4}
  \Delta x^4 \,\Delta t^{-1} &\mapsto 3 \  \lceil{\Delta x^2}\rfloor^2 \,\Delta t^{-1}
  \; , \\
  \label{eq:Ito6}
  \Delta x^6 \,\Delta t^{-2} &\mapsto 15 \ \lceil{\Delta x^2}\rfloor^3 \,\Delta t^{-2}
  \; . 
\end{align}
Note that, as discussed in Ref.~\cite{Cugliandolo-Lecomte17a}, the substitution rule~\eref{eq:Ito3} \emph{cannot} be used inside the exponential of the infinitesimal propagator; indeed, since $\Delta x^3 \,\Delta t^{-1}=O(\Delta t^{1/2})$ one has $\ee^{h(x) \Delta x^3 \,\Delta t^{-1}}=1+h(x)\Delta x^3 \,\Delta t^{-1}+\tfrac 12 [h(x)\Delta x^3 \,\Delta t^{-1}]^2+O(\Delta t^{3/2})$
and the second term of this expansion would be wrong if one had first applied the rule~\eref{eq:Ito3} and then expanded.
This is the trivial but shrouded reason why the procedure exposed in Appendix~\ref{sec:ch-var-discrete} has to be performed by expanding the terms of order $\Delta t^{>0}$ outside of the exponential of the infinitesimal propagator of \eref{eq:expansionpolynomialPxfromPU}.
This reflects the fact, known to mathematicians, that the validity of the continuous-time chain rule  is relatively weak, even in the Stratonovich discretization: it cannot be manipulated without care by, for instance, taking its square and exponentiating it --~as one would do 
by naively using it in the Onsager--Machlup action.
For further discussion on this subject, see Ref.~\cite{Cugliandolo-Lecomte17a}.

%_______________________________________________________________________________________________________
%_______________________________________________________________________________________________________

\section*{References}

\addcontentsline{toc}{section}{References}
\bibliographystyle{plain_url}

\bibliography{path}

\end{document}